\documentclass[%
 aip,
apl,%
 amsmath,amssymb,
 reprint,%
bibliography,
]{revtex4-1}

\usepackage{graphicx}
\usepackage{dcolumn}
\usepackage{bm}
\usepackage{epstopdf}

\begin{document}


\title{A microfabricated optically-pumped magnetic gradiometer}

\author{D. Sheng}~\thanks{Current Address: Department of Precision Machinery and Precision Instrumentation, University of Science and Technology of China, Hefei 230027, China}
\author{A. R. Perry}
\author{S. P. Krzyzewski}
 \affiliation{Time and Frequency Division, National Institute of Standards and Technology, 325 Broadway, Boulder, Colorado 80305, USA}
\affiliation{University of Colorado, Boulder, Colorado 80309, USA}
\author{S. Geller}
 \affiliation{Time and Frequency Division, National Institute of Standards and Technology, 325 Broadway, Boulder, Colorado 80305, USA}
\author{J. Kitching}%
 \affiliation{Time and Frequency Division, National Institute of Standards and Technology, 325 Broadway, Boulder, Colorado 80305, USA}
\author{S. Knappe}
 \affiliation{Time and Frequency Division, National Institute of Standards and Technology, 325 Broadway, Boulder, Colorado 80305, USA}
\affiliation{University of Colorado, Boulder, Colorado 80309, USA}

\date{\today}

\begin{abstract}
We report on the development of a microfabricated atomic magnetic gradiometer based on optical spectroscopy
of alkali atoms in the vapor phase. The gradiometer, which operates in the spin-exchange relaxation free regime, has a length of 60 mm and cross sectional diameter of 12 mm, and consists of two chip-scale atomic magnetometers which are interrogated by a common laser light. The sensor can measure differences in magnetic fields,
over a 20 mm baseline, of 10 fT/Hz$^{1/2}$ at frequencies above 20 Hz. The maximum rejection of magnetic field noise is 1000 at 10 Hz. By use of a set of compensation coils wrapped around the sensor, we also measure the sensor sensitivity at several external bias field strengths up to 150 mG. This device is useful for applications that require both sensitive gradient field information and high common-mode noise cancellation.
\end{abstract}

\maketitle

Optically-pumped magnetometers operate with a net spin polarization, and the spin undergoes Larmor precession in an external magnetic field~\cite{budker13}. This spin rotation modifies the light-atom interaction, and a magnetic field dependent signal appears in either the absorption or phase shift of a probe beam~\cite{budker07}. Two main research directions in this field have been to improve the sensor sensitivity on the one hand and the size, weight, power consumption and spatial resolution on the other hand. Subfemtotesla magnetic field sensitivity has been achieved in the spin-exchange relaxation free (SERF) regime~\cite{kominis03,dang2010}, and with radio-frequency~\cite{lee06}, and scalar~\cite{sheng13} atomic magnetometers. Micrometer and nanometer scale spatial resolutions have also been realized in systems such as Bose-Einstein condensates~\cite{vengalattore07,eto14} and nitrogen-vacancy centers~\cite{maletinsky12}, respectively. The combination of these two research directions has led to compact cavity-assisted table-top experiments~\cite{Li12,sheng13,jensen14,clevenson15,vasilakis15} and highly sensitive miniaturized devices~\cite{mhaskar12}. These devices, making use of millimeter-scale cells, are small, power efficient, and broadly useful in sensing applications that require portability or battery operations. They are also useful for biomagnetic imaging, where the target distance is on the millimeter to centimeter scale. Such applications include human magnetoencephalography~\cite{sander12} and magnetocardiography~\cite{knappe10}. For better target field spatial information and common noise cancellation in these applications, it requires gradiometer sensors.  Here we report on the development of a miniaturized magnetic gradiometer based on chip-scale atomic magnetometers.

\begin{figure}
\includegraphics[width=3.1in]{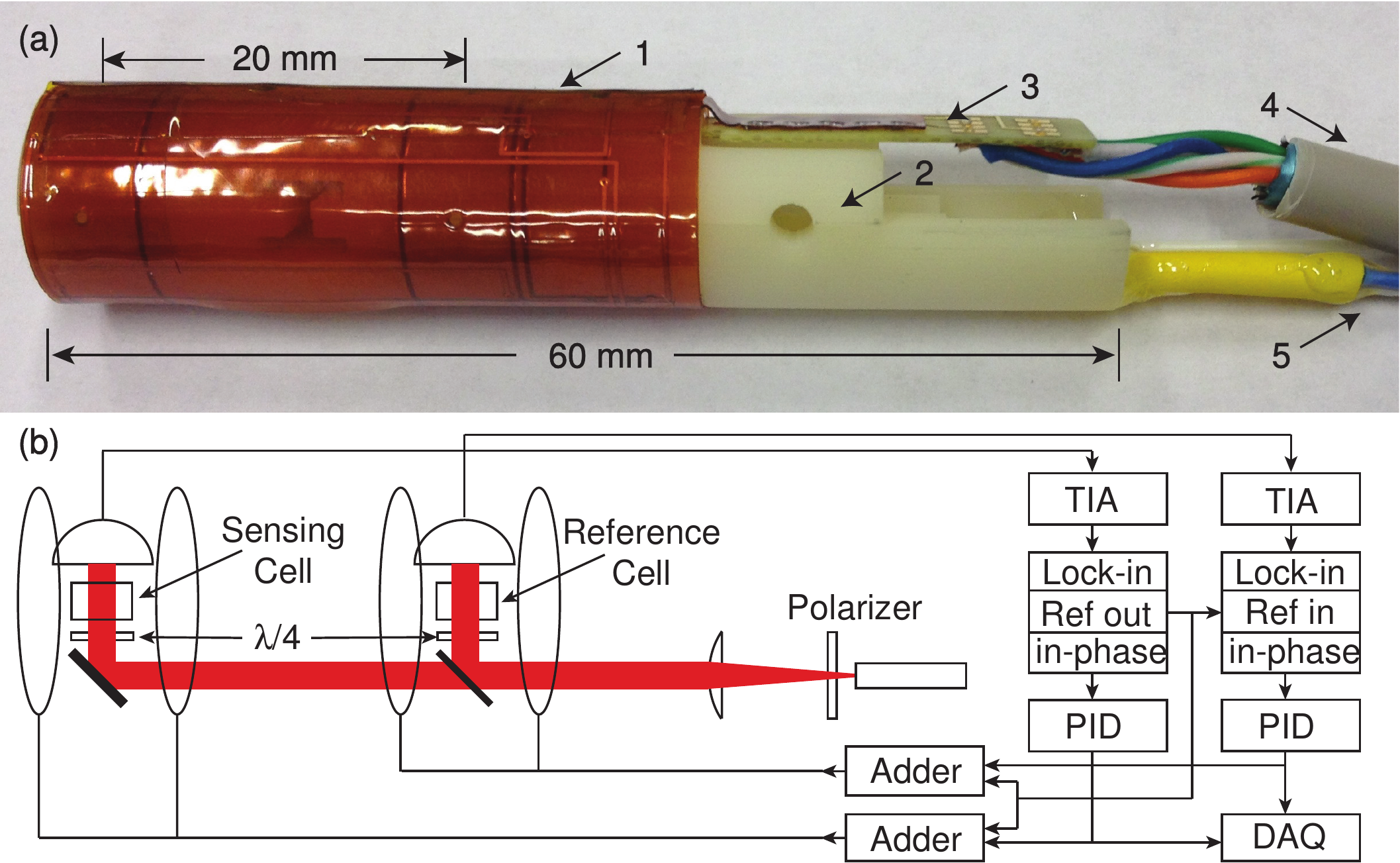}
\caption{\label{fig:setup}(Color online) (a) Photograph of the magnetic gradiometer. 1. Kapton flexible coil, 2. optical bench, 3. PCB, 4. CAT5e cable, 5. optical fibers. (b) Plot of the optical paths inside the sensor and electrical signal processes (heating beams not shown). TIA: trans-impedance amplifier, PID:  proportional-integral-derivative controller.}
\end{figure}

NIST chip-scale magnetometers~\cite{schwindt04,mhaskar12} use microfabricated vapor cells, and operate with a single laser beam for both pumping and probing the atomic polarization. The high magnetic field sensitivity can result from a combination of the SERF mechanism~\cite{happer73,allred02} and a zero field level crossing resonance~\cite{cohen70}.  We build the gradiometer by placing two such cells on the same optical bench separated by 20~mm as shown in Fig.~\ref{fig:setup}(a). This geometry allows both cells to share the same laser  source. As a result, the optical noise is largely common to both magnetometers. Compared to other gradiometers, which use two separate magnetometer devices~\cite{shah13} or a single large cell~\cite{johnson10}, the sensor in this paper maintains both a high common-mode rejection ratio (CMRR) and a small cell size. 

The optical bench is a cylinder with a length of 60 mm and a diameter of 12 mm, fabricated using high-resolution stereolithography. The bench is fabricated in two halves separated by a plane in the cylinder axis. One half contains the photodetectors, while the other half contains all the optics.  The cells are located 4~mm and 24~mm from the end of the sensor, respectively, which allows the sensing cell to be placed in close proximity to the magnetic field source. 

There are three separate optical beams inside the sensor. One beam at 795 nm is used for the optical pumping, which is split into two beams, one for each cell. Two separate beams at 1550 nm are used for independently heating the cells~\cite{preusser09}. Fig.~\ref{fig:setup}(b) shows the pumping beam paths inside the sensor. This pumping beam is generated by a distributed feedback diode laser, and sent to the bench through a 6 m long polarization maintaining fiber. A 0.2 mm thick polarizer at the end of the fiber cleans up the beam polarization.  The diverging beam goes through a lens anti-reflection coated for wavelengths from 700 nm to 1600 nm, and the output collimated beam diameter is about 2 mm. A dichroic mirror is placed next to the reference cell, acting as a 50/50 beam splitter for the pumping beam, and a reflector for the reference cell heating beam. A gold coated mirror under the sensing cell reflects both its pumping and heating beams. Thin quarter-wave plates circularly polarize the pumping beams, which are detected after the cells by two silicon photodiode chips with an active area of 3.5 mm$^2$. The photodiode chips are attached to a printed circuit board (PCB), using conductive glue for the cathodes and wire bonding for the anodes.  A printed Kapton flexible coil system is wrapped around the optical bench. This flex coil system provides two pairs of circular Helmholtz coils for magnetic fields in the main sensitive direction along the cylinder axis of the sensor, and two pairs of saddle coils for common magnetic fields in the two perpendicular directions. The flex coils are soldered to the PCB, so that all electrical controls exit the sensor through the PCB. A 6 m long CAT5e cable transfers the electrical signals to the control electronics.

The chip-scale cells used in this sensor are made using anodic bonding between silicon and Pyrex glass.  The cell's outer size is 4$\times$4$\times$2.7 mm$^3$ with an interior size of 3$\times$3$\times$2 mm$^3$. The cells were filled with roughly 1 amg of N$_2$ gas, and a droplet of $^{87}$Rb. If cells in a gradiometer are filled with different gas pressures, Rb atoms experience different shifts of the optical absorption line. This reduces the laser noise cancellation in the gradiometer, because the same pumping beam frequency noise coverts to different amplitude noise of the transmitted beam. To minimize this effect, we pick two cells  which differ in the pressure shifts of the Rb $D1$ line by less than 0.3 GHz, which corresponds to a N$_2$ pressure difference less than 0.04 amg. We attach pieces of 0.2 and 0.6 mm thick color glass filter to the front and back of the cell to heat uniformly the cells through the absorption of 1550 nm light~\cite{mhaskar12}. 

The cells were suspended on a Kapton web inside a vacuum package~\cite{knappe16} with dimensions of 8.5$\times$8$\times$5 mm$^3$. The heat loss of a vacuum-packed cell is dominated by the Stefan-Boltzmann law,
\begin{equation}
\label{eq:sblaw}
P=A\sigma(T_c^4-T_e^4),
\end{equation}
where A is the surface area of the cell with colored glass attached, $T_c$ ($T_e$) is the cell (environment) temperature, and $\sigma$ is the Stefan-Boltzmann constant. At $T_e=20~^\circ$C, Eq.~(\ref{eq:sblaw}) predicts that it takes 115 mW power to heat the cell to 150 $^\circ$C. Experimentally a heat beam power of 150 mW was required to reach the same temperature, which confirms that the heat loss of our sensor is dominated by the black-body radiation of the cell.  Eq.~(\ref{eq:sblaw}) also predicts that, for fixed heating power, $T_c$ changes by 7.5 $^\circ$C when $T_e$ changes by 20 $^\circ$C. We confirmed this dependence by adding an oven around the sensor and monitoring the change in beam transmission. Since the SERF regime covers a broad temperature range, a change in $T_c$ results mainly in a change of the pumping beam transmission on a long time scale. This effect can be diminished by stabilizing the cell
temperature through monitoring the dc level of the photodiode output and feeding back to the heater power.

A SERF magnetometer requires not only a high cell temperature for a large atomic density, but also a low magnetic  field environment. In this case, a stable atomic polarization $P=2\langle{S_p}\rangle$ builds up, where $S_p$ is the electron spin along the pumping beam direction. A magnetic field modulation of frequency $\omega$, which is much larger than the Zeeman frequency and optical pumping rate, perpendicular to the pumping beam direction results in modulated polarization components at harmonics of $\omega$, with the first harmonic component as~\cite{cohen70,shah09},
\begin{eqnarray}
\label{Eq:mod}
P(\omega)=&&\frac{\gamma_eB_{t,0}R_p\sin{\omega{t}}}{(R_{r}+R_p)^2+(\gamma_eB_{t,0})^2}\times\nonumber\\
&&J_0(\frac{\gamma_eB_{t,m}}{Q(P)\omega})J_1(\frac{\gamma_eB_{t,m}}{Q(P)\omega}),
\end{eqnarray}
where $J$ is Bessel function of the first kind, $\gamma_e$ is the electron gyromagnetic ratio, $Q(P)$ is the nuclear spin slow-down factor, $B_{t,0}$ and $B_{t,m}$ are the offset and modulation fields in the transverse direction, respectively, $R_p$ is the pumping rate, and $R_r$ is the spin relaxation rate which is dominated by the spin-destruction relaxation rate in the SERF regime. Because the pumping beam absorption is related to the atomic spin polarization, there are corresponding modulations of the beam transmission $I$ at harmonics of $\omega$. Its first harmonic component $I(\omega)$ shows a similar dispersive relation with the offset field as $P(\omega)$. In the experiment, we use the reference output of a lock-in amplifier to provide the modulation field, which also demodulates the transmitted beam signal at frequency $\omega$. We pass the in-phase output signal to a PID controller, feed back the PID output to the Helmholtz coils on the sensor, and read out the field noise from the feedback signal. The conversion from a feedback signal to a magnetic field depends on the coil field calibration only. Therefore, it is more robust to changes in sensor parameters than in the case of open-loop operations~\cite{mhaskar12}, which leads to a better CMRR. 

\begin{figure}
\includegraphics[width=3.1in]{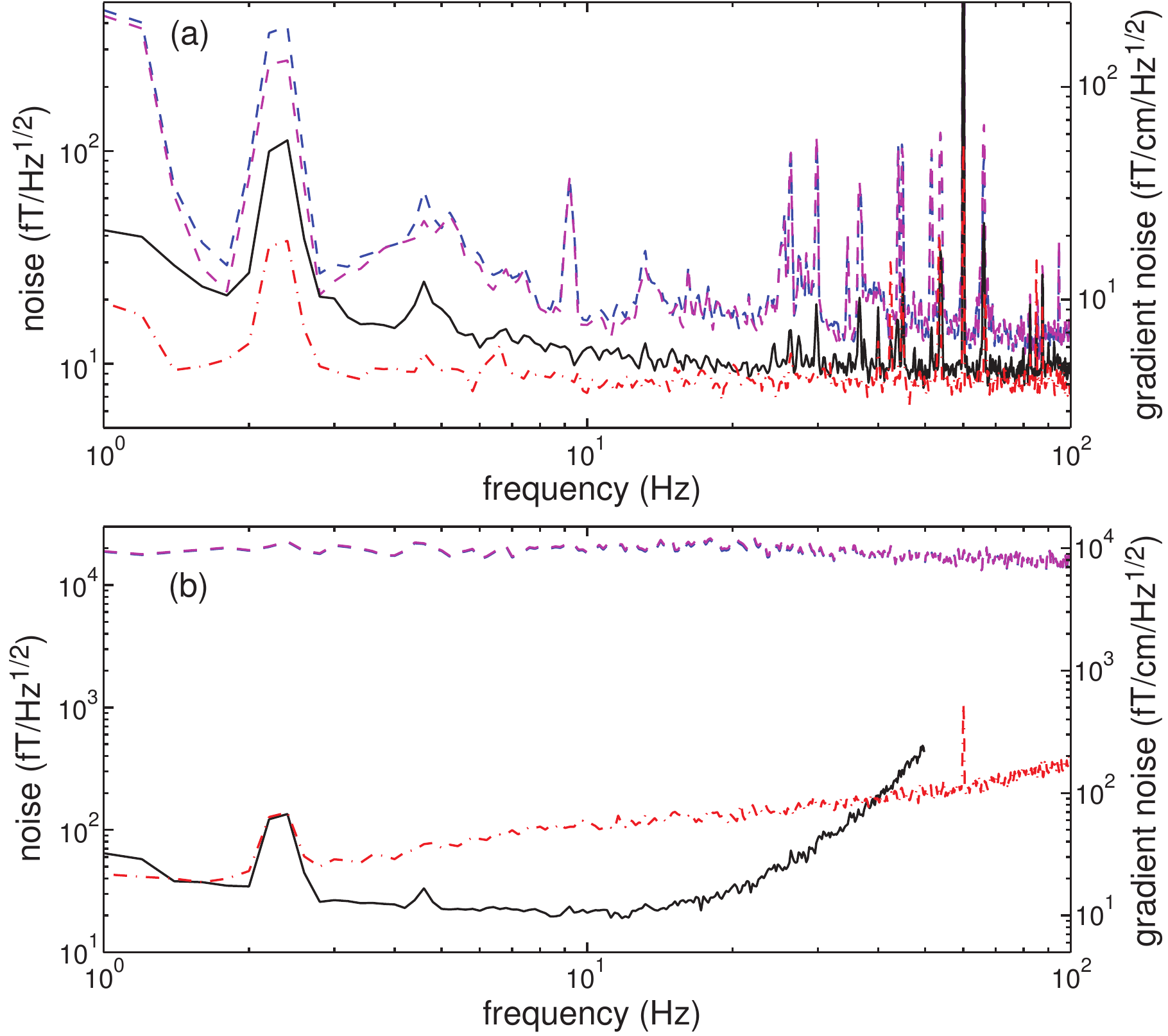}
\caption{\label{fig:cmrr}(Color online) (a) Plot of the magnetic field and gradient field sensitivity of the sensor. The blue (magenta) dash line corresponds to the reference (sensing) cell result,  the black solid line is the noise of the difference between them. The red dash-dot line is the subtraction of the lock-in amplifier quadrature outputs. (b) Plot of the gradiometer CMRR with different PID settings. The blue (magenta) dash line corresponds to the reference (sensing) cell result, the red dash-dot (black solid) line corresponds to the gradiometer result using a single (double) integrator in the PID controller.}
\end{figure}

The gradiometer sensor was tested in a three-layer magnetically-shielded room. The sensor works in a free-running mode, without optical power, wavelength, or temperature control.  Fig.~\ref{fig:cmrr}(a) shows the sensor performance with 1.2 mW of power coupled into the pumping beam fiber, 150 mW of power coupled into each heating beam fiber, and 100 nT modulation field at 1.79 kHz. Each magnetometer has an open-loop bandwidth of 130 Hz, and a magnetic field sensitivity better than 20 fT/Hz$^{1/2}$ at frequencies above 10 Hz. An independent table-top experiment~\cite{krzyzewskiprep} with similar cells and better control of the pumping-beam noise shows that it is possible to improve the single magnetometer sensitivity to better than 10 fT/Hz$^{1/2}$. The cross-talk effect of the Helmholtz coils of one cell onto the other cell is less than 4 \%. We acquire the gradiometer results by subtracting the two magnetometer results. The gradiometer sensitivity is around 5 fT/cm/Hz$^{1/2}$ at frequencies above 20 Hz. This agrees with the photon-noise-limited sensitivity in the open-loop operation, which we find by subtracting the lock-in amplifier quadrature outputs~\cite{mhaskar12}. We also find that the gradiometer sensitivity degrades by less than 10~\%, when the pumping beam frequency is tuned within a 5 GHz range. This makes it possible to run multiple gradiometers with different buffer-gas pressure shifts using a single pumping-beam source, as long as the two cells inside each gradiometer have similar gas pressures.

The common-mode noise rejection ratio is an important parameter to characterize the gradiometer. A large CMRR implies that the gradiometer could work in a noisy environment, which relaxes the requirements on the test room shielding factor or the instrument noise of nulling fields. To measure the CMRR, we place the gradiometer at the center of a three-dimensional rectangular Helmholtz coil system with side lengths of about 1 m. We pass white noise  through a low pass filter with a bandwidth of 150 Hz, and connect the output to the large Helmholtz coils along the sensitive gradiometer direction.  Fig.~\ref{fig:cmrr}(b) shows that, with a single integrator in the PID controller, the sensor CMRR is about 500 at 1 Hz, and larger than 100 at frequencies up to 50 Hz. By adding a second integrator in the PID controller, we achieve better noise cancellation in the feedback loop at the cost of bandwidth reduction. In this case, the gradiometer shows a CMRR up to 1000 at frequencies near 10 Hz, where the best CMRR frequency is related to the time constant of the second integrator. This is the best CMRR reported for magnetic gradiometers in the SERF regime~\cite{Note1}.

The suppression of spin-exchange relaxation in the SERF regime limits the sensor dynamic range to 10 nT. If the bias field is beyond this dynamic range at the time of turning on the gradiometer, its control system is not able to provide the correct feedback current to null the offset field. A way around this problem is to add an additional compensation coil system and a sensor with a larger dynamic range to first bring the bias field to within the dynamic range of atomic sensor~\cite{seltzer08}.  To implement this scheme in our gradiometer sensor while keeping the sensor size unchanged requires a second small and stable magnetic sensor. An example of such a sensor is the Honeywell HMC1053 three-axis magnetoresistive device~\cite{Note2}. We have tested this sensor in the polarity switching mode,  and confirmed that the readout is stable within 10 nT over 30 hours with a proper thermal isolation. 

\begin{figure}
\includegraphics[width=3.1in]{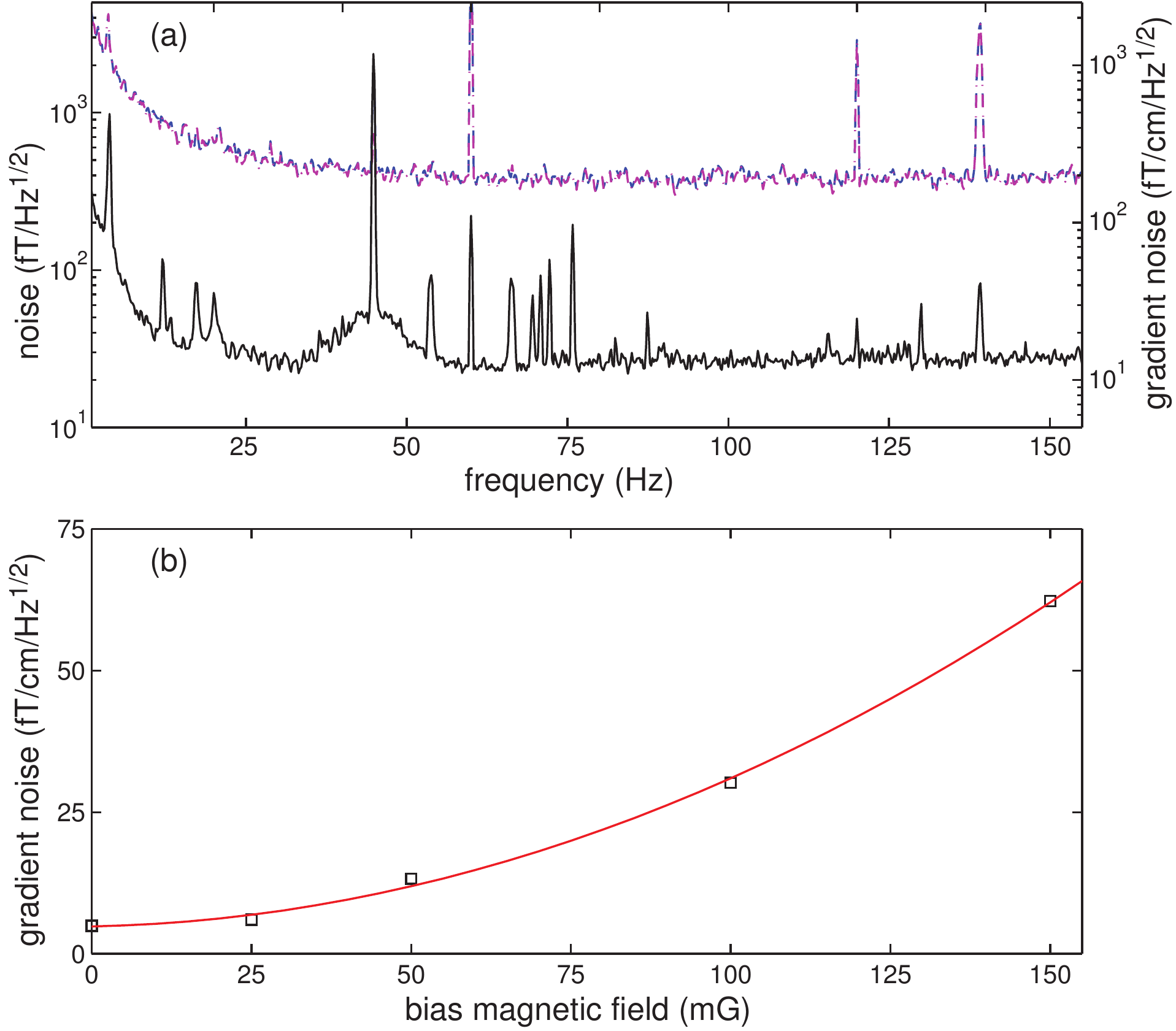}
\caption{\label{fig:bias}(Color online). (a) Plot of the magnetic field and gradient field sensitivity of the sensor at a bias field of 50 mG. The blue (magenta) dash line corresponds to the reference (sensing) cell result, and the black solid line is the difference of the two magnetometer results. (b) The averaged gradient magnetic field sensitivity around 100 Hz at different bias fields. Black empty box (red solid line) is the experimental (fitting) result.}
\end{figure}

The compensation coils are also required to be small to keep the whole system compact. We wrap a pair of Helmholtz and anti-Helmholtz coils for each cell on a slotted cylinder attached to the sensor outer surface as compensation coils.  Compared to references which used a large compensation coil system~\cite{seltzer04,belfi10}, the small compensation coils compatible with our sensor could generate significant field gradients. To test the limitation in this small system,  we use the 1 m size Helmholtz coil system mentioned previously to generate an offset field, and measure the gradiometer sensitivity after zeroing the offset field using the sensor compensation coils. Fig.~\ref{fig:bias}(a) shows the result at a 50 mG bias field. The sensitivity of each magnetometer is dominated by the  noise from the bias field current source~\cite{Note3}. The gradiometer largely cancels this common noise,  and its sensitivity is limited by  the field gradients. To compare the gradiometer sensitivity at different bias fields, we take the average gradiometer sensitivity from 90 Hz to 110 Hz, and plot the results in Fig.~\ref{fig:bias}(b). The gradiometer noise level increases quadratically with the bias field strength, which agrees with the expected scaling due to the gradient field~\cite{torrey56}. With this scaling,  the sensitivity of this gradiometer sensor is expected to be around  0.5~pT/cm/Hz$^{1/2}$ at the earth field amplitude, comparable with several previous measurements using a larger system \cite{seltzer04,belfi10}. By adding more field gradient cancellation coils to the sensor, it would be possible to reduce the gradient field by another factor of three, this would reduce the noise level at earth field to be less than 100 fT/cm/Hz$^{1/2}$, comparable to a recent result using a scalar gradiometer with a large compensation coil system~\cite{bevilacqua16}. This could potentially lead to a compact biomagnetic imaging system in an unshielded environment.

In summary, we have developed a miniaturized atomic magnetic gradiometer using vacuum-packaged chip-scale cells. This gradiometer works with a single laser beam, and operates in the SERF regime. Its heating power consumption is dominated by the black-body radiation from the cells. This gradiometer shows a gradient field sensitivity of 5 fT/cm/Hz$^{1/2}$ with a 20 mm baseline, and a maximum CMRR of 1000 within the sensor bandwidth. By adding a set of coils around the sensor to cancel the offset magnetic field, we test the sensor performance at external bias fields up to 150 mG. We are working to improve the sensitivity of this system by one order of magnitude in an unshielded environment. This device is useful for applications that require both sensitive gradient field information and high CMRR.

Funding from NIH and NIST is acknowledged. The authors thank Susan Schima for fabrication support. This work is a contribution of NIST, an agency of U.S. Government, and is not subject to copyright.

%


\begin{thebibliography}{34}%
\makeatletter
\providecommand \@ifxundefined [1]{%
 \@ifx{#1\undefined}
}%
\providecommand \@ifnum [1]{%
 \ifnum #1\expandafter \@firstoftwo
 \else \expandafter \@secondoftwo
 \fi
}%
\providecommand \@ifx [1]{%
 \ifx #1\expandafter \@firstoftwo
 \else \expandafter \@secondoftwo
 \fi
}%
\providecommand \natexlab [1]{#1}%
\providecommand \enquote  [1]{``#1''}%
\providecommand \bibnamefont  [1]{#1}%
\providecommand \bibfnamefont [1]{#1}%
\providecommand \citenamefont [1]{#1}%
\providecommand \href@noop [0]{\@secondoftwo}%
\providecommand \href [0]{\begingroup \@sanitize@url \@href}%
\providecommand \@href[1]{\@@startlink{#1}\@@href}%
\providecommand \@@href[1]{\endgroup#1\@@endlink}%
\providecommand \@sanitize@url [0]{\catcode `\\12\catcode `\$12\catcode
  `\&12\catcode `\#12\catcode `\^12\catcode `\_12\catcode `\%12\relax}%
\providecommand \@@startlink[1]{}%
\providecommand \@@endlink[0]{}%
\providecommand \url  [0]{\begingroup\@sanitize@url \@url }%
\providecommand \@url [1]{\endgroup\@href {#1}{\urlprefix }}%
\providecommand \urlprefix  [0]{URL }%
\providecommand \Eprint [0]{\href }%
\providecommand \doibase [0]{http://dx.doi.org/}%
\providecommand \selectlanguage [0]{\@gobble}%
\providecommand \bibinfo  [0]{\@secondoftwo}%
\providecommand \bibfield  [0]{\@secondoftwo}%
\providecommand \translation [1]{[#1]}%
\providecommand \BibitemOpen [0]{}%
\providecommand \bibitemStop [0]{}%
\providecommand \bibitemNoStop [0]{.\EOS\space}%
\providecommand \EOS [0]{\spacefactor3000\relax}%
\providecommand \BibitemShut  [1]{\csname bibitem#1\endcsname}%
\let\auto@bib@innerbib\@empty
\bibitem [{\citenamefont {Budker}\ and\ \citenamefont
  {Kimball}(2013)}]{budker13}%
  \BibitemOpen
  \bibfield  {author} {\bibinfo {author} {\bibfnamefont {D.}~\bibnamefont
  {Budker}}\ and\ \bibinfo {author} {\bibfnamefont {D.~F.~J.}\ \bibnamefont
  {Kimball}},\ }\href@noop {} {\emph {\bibinfo {title} {Optical
  magnetometry}}}\ (\bibinfo  {publisher} {Cambridge University Press},\
  \bibinfo {year} {2013})\BibitemShut {NoStop}%
\bibitem [{\citenamefont {Budker}\ and\ \citenamefont
  {Romalis}(2007)}]{budker07}%
  \BibitemOpen
  \bibfield  {author} {\bibinfo {author} {\bibfnamefont {D.}~\bibnamefont
  {Budker}}\ and\ \bibinfo {author} {\bibfnamefont {M.}~\bibnamefont
  {Romalis}},\ }\href@noop {} {\bibfield  {journal} {\bibinfo  {journal}
  {Nature Phys.}\ }\textbf {\bibinfo {volume} {3}},\ \bibinfo {pages} {227}
  (\bibinfo {year} {2007})}\BibitemShut {NoStop}%
\bibitem [{\citenamefont {Kominis}\ \emph {et~al.}(2003)\citenamefont
  {Kominis}, \citenamefont {Kornack}, \citenamefont {Allred},\ and\
  \citenamefont {Romalis}}]{kominis03}%
  \BibitemOpen
  \bibfield  {author} {\bibinfo {author} {\bibfnamefont {I.~K.}\ \bibnamefont
  {Kominis}}, \bibinfo {author} {\bibfnamefont {T.~W.}\ \bibnamefont
  {Kornack}}, \bibinfo {author} {\bibfnamefont {J.~C.}\ \bibnamefont {Allred}},
  \ and\ \bibinfo {author} {\bibfnamefont {M.~V.}\ \bibnamefont {Romalis}},\
  }\href@noop {} {\bibfield  {journal} {\bibinfo  {journal} {Nature}\ }\textbf
  {\bibinfo {volume} {422}},\ \bibinfo {pages} {596} (\bibinfo {year}
  {2003})}\BibitemShut {NoStop}%
\bibitem [{\citenamefont {Dang}, \citenamefont {Maloof},\ and\ \citenamefont
  {Romalis}(2010)}]{dang2010}%
  \BibitemOpen
  \bibfield  {author} {\bibinfo {author} {\bibfnamefont {H.~B.}\ \bibnamefont
  {Dang}}, \bibinfo {author} {\bibfnamefont {A.~C.}\ \bibnamefont {Maloof}}, \
  and\ \bibinfo {author} {\bibfnamefont {M.~V.}\ \bibnamefont {Romalis}},\
  }\href {\doibase http://dx.doi.org/10.1063/1.3491215} {\bibfield  {journal}
  {\bibinfo  {journal} {Appl. Phys. Lett.}\ }\textbf {\bibinfo {volume} {97}},\
  \bibinfo {eid} {151110} (\bibinfo {year} {2010})}\BibitemShut {NoStop}%
\bibitem [{\citenamefont {Lee}\ \emph {et~al.}(2006)\citenamefont {Lee},
  \citenamefont {Sauer}, \citenamefont {Seltzer}, \citenamefont {Alem},\ and\
  \citenamefont {Romalis}}]{lee06}%
  \BibitemOpen
  \bibfield  {author} {\bibinfo {author} {\bibfnamefont {S.-K.}\ \bibnamefont
  {Lee}}, \bibinfo {author} {\bibfnamefont {K.~L.}\ \bibnamefont {Sauer}},
  \bibinfo {author} {\bibfnamefont {S.~J.}\ \bibnamefont {Seltzer}}, \bibinfo
  {author} {\bibfnamefont {O.}~\bibnamefont {Alem}}, \ and\ \bibinfo {author}
  {\bibfnamefont {M.~V.}\ \bibnamefont {Romalis}},\ }\href@noop {} {\bibfield
  {journal} {\bibinfo  {journal} {Appl. Phys. Lett.}\ }\textbf {\bibinfo
  {volume} {89}},\ \bibinfo {pages} {214106} (\bibinfo {year}
  {2006})}\BibitemShut {NoStop}%
\bibitem [{\citenamefont {Sheng}\ \emph {et~al.}(2013)\citenamefont {Sheng},
  \citenamefont {Li}, \citenamefont {Dural},\ and\ \citenamefont
  {Romalis}}]{sheng13}%
  \BibitemOpen
  \bibfield  {author} {\bibinfo {author} {\bibfnamefont {D.}~\bibnamefont
  {Sheng}}, \bibinfo {author} {\bibfnamefont {S.}~\bibnamefont {Li}}, \bibinfo
  {author} {\bibfnamefont {N.}~\bibnamefont {Dural}}, \ and\ \bibinfo {author}
  {\bibfnamefont {M.~V.}\ \bibnamefont {Romalis}},\ }\href {\doibase
  10.1103/PhysRevLett.110.160802} {\bibfield  {journal} {\bibinfo  {journal}
  {Phys. Rev. Lett.}\ }\textbf {\bibinfo {volume} {110}},\ \bibinfo {pages}
  {160802} (\bibinfo {year} {2013})}\BibitemShut {NoStop}%
\bibitem [{\citenamefont {Vengalattore}\ \emph {et~al.}(2007)\citenamefont
  {Vengalattore}, \citenamefont {Higbie}, \citenamefont {Leslie}, \citenamefont
  {Guzman}, \citenamefont {Sadler},\ and\ \citenamefont
  {Stamper-Kurn}}]{vengalattore07}%
  \BibitemOpen
  \bibfield  {author} {\bibinfo {author} {\bibfnamefont {M.}~\bibnamefont
  {Vengalattore}}, \bibinfo {author} {\bibfnamefont {J.}~\bibnamefont
  {Higbie}}, \bibinfo {author} {\bibfnamefont {S.}~\bibnamefont {Leslie}},
  \bibinfo {author} {\bibfnamefont {J.}~\bibnamefont {Guzman}}, \bibinfo
  {author} {\bibfnamefont {L.}~\bibnamefont {Sadler}}, \ and\ \bibinfo {author}
  {\bibfnamefont {D.}~\bibnamefont {Stamper-Kurn}},\ }\href@noop {} {\bibfield
  {journal} {\bibinfo  {journal} {Phys. Rev. Lett.}\ }\textbf {\bibinfo
  {volume} {98}},\ \bibinfo {pages} {200801} (\bibinfo {year}
  {2007})}\BibitemShut {NoStop}%
\bibitem [{\citenamefont {Eto}, \citenamefont {Saito},\ and\ \citenamefont
  {Hirano}(2014)}]{eto14}%
  \BibitemOpen
  \bibfield  {author} {\bibinfo {author} {\bibfnamefont {Y.}~\bibnamefont
  {Eto}}, \bibinfo {author} {\bibfnamefont {H.}~\bibnamefont {Saito}}, \ and\
  \bibinfo {author} {\bibfnamefont {T.}~\bibnamefont {Hirano}},\ }\href@noop {}
  {\bibfield  {journal} {\bibinfo  {journal} {Phys. Rev. Lett.}\ }\textbf
  {\bibinfo {volume} {112}},\ \bibinfo {pages} {185301} (\bibinfo {year}
  {2014})}\BibitemShut {NoStop}%
\bibitem [{\citenamefont {Maletinsky}\ \emph {et~al.}(2012)\citenamefont
  {Maletinsky}, \citenamefont {Hong}, \citenamefont {Grinolds}, \citenamefont
  {Hausmann}, \citenamefont {Lukin}, \citenamefont {Walsworth}, \citenamefont
  {Loncar},\ and\ \citenamefont {Yacoby}}]{maletinsky12}%
  \BibitemOpen
  \bibfield  {author} {\bibinfo {author} {\bibfnamefont {P.}~\bibnamefont
  {Maletinsky}}, \bibinfo {author} {\bibfnamefont {S.}~\bibnamefont {Hong}},
  \bibinfo {author} {\bibfnamefont {M.~S.}\ \bibnamefont {Grinolds}}, \bibinfo
  {author} {\bibfnamefont {B.}~\bibnamefont {Hausmann}}, \bibinfo {author}
  {\bibfnamefont {M.~D.}\ \bibnamefont {Lukin}}, \bibinfo {author}
  {\bibfnamefont {R.~L.}\ \bibnamefont {Walsworth}}, \bibinfo {author}
  {\bibfnamefont {M.}~\bibnamefont {Loncar}}, \ and\ \bibinfo {author}
  {\bibfnamefont {A.}~\bibnamefont {Yacoby}},\ }\href@noop {} {\bibfield
  {journal} {\bibinfo  {journal} {Nat. Nanotechnology}\ }\textbf {\bibinfo
  {volume} {7}},\ \bibinfo {pages} {320} (\bibinfo {year} {2012})}\BibitemShut
  {NoStop}%
\bibitem [{\citenamefont {Li}\ \emph {et~al.}(2011)\citenamefont {Li},
  \citenamefont {Vachaspati}, \citenamefont {Sheng}, \citenamefont {Dural},\
  and\ \citenamefont {Romalis}}]{Li12}%
  \BibitemOpen
  \bibfield  {author} {\bibinfo {author} {\bibfnamefont {S.}~\bibnamefont
  {Li}}, \bibinfo {author} {\bibfnamefont {P.}~\bibnamefont {Vachaspati}},
  \bibinfo {author} {\bibfnamefont {D.}~\bibnamefont {Sheng}}, \bibinfo
  {author} {\bibfnamefont {N.}~\bibnamefont {Dural}}, \ and\ \bibinfo {author}
  {\bibfnamefont {M.~V.}\ \bibnamefont {Romalis}},\ }\href@noop {} {\bibfield
  {journal} {\bibinfo  {journal} {Phys. Rev. A}\ }\textbf {\bibinfo {volume}
  {84}},\ \bibinfo {pages} {061403} (\bibinfo {year} {2011})}\BibitemShut
  {NoStop}%
\bibitem [{\citenamefont {Jensen}\ \emph {et~al.}(2014)\citenamefont {Jensen},
  \citenamefont {Leefer}, \citenamefont {Jarmola}, \citenamefont {Dumeige},
  \citenamefont {Acosta}, \citenamefont {Kehayias}, \citenamefont {Patton},\
  and\ \citenamefont {Budker}}]{jensen14}%
  \BibitemOpen
  \bibfield  {author} {\bibinfo {author} {\bibfnamefont {K.}~\bibnamefont
  {Jensen}}, \bibinfo {author} {\bibfnamefont {N.}~\bibnamefont {Leefer}},
  \bibinfo {author} {\bibfnamefont {A.}~\bibnamefont {Jarmola}}, \bibinfo
  {author} {\bibfnamefont {Y.}~\bibnamefont {Dumeige}}, \bibinfo {author}
  {\bibfnamefont {V.~M.}\ \bibnamefont {Acosta}}, \bibinfo {author}
  {\bibfnamefont {P.}~\bibnamefont {Kehayias}}, \bibinfo {author}
  {\bibfnamefont {B.}~\bibnamefont {Patton}}, \ and\ \bibinfo {author}
  {\bibfnamefont {D.}~\bibnamefont {Budker}},\ }\href@noop {} {\bibfield
  {journal} {\bibinfo  {journal} {Phys. Rev. Lett.}\ }\textbf {\bibinfo
  {volume} {112}},\ \bibinfo {pages} {160802} (\bibinfo {year}
  {2014})}\BibitemShut {NoStop}%
\bibitem [{\citenamefont {Clevenson}\ \emph {et~al.}(2015)\citenamefont
  {Clevenson}, \citenamefont {Trusheim}, \citenamefont {Teale}, \citenamefont
  {Schr{\"o}der}, \citenamefont {Braje},\ and\ \citenamefont
  {Englund}}]{clevenson15}%
  \BibitemOpen
  \bibfield  {author} {\bibinfo {author} {\bibfnamefont {H.}~\bibnamefont
  {Clevenson}}, \bibinfo {author} {\bibfnamefont {M.~E.}\ \bibnamefont
  {Trusheim}}, \bibinfo {author} {\bibfnamefont {C.}~\bibnamefont {Teale}},
  \bibinfo {author} {\bibfnamefont {T.}~\bibnamefont {Schr{\"o}der}}, \bibinfo
  {author} {\bibfnamefont {D.}~\bibnamefont {Braje}}, \ and\ \bibinfo {author}
  {\bibfnamefont {D.}~\bibnamefont {Englund}},\ }\href@noop {} {\bibfield
  {journal} {\bibinfo  {journal} {Nat. Phys.}\ }\textbf {\bibinfo {volume}
  {11}},\ \bibinfo {pages} {393} (\bibinfo {year} {2015})}\BibitemShut
  {NoStop}%
\bibitem [{\citenamefont {Vasilakis}\ \emph {et~al.}(2015)\citenamefont
  {Vasilakis}, \citenamefont {Shen}, \citenamefont {Jensen}, \citenamefont
  {Balabas}, \citenamefont {Salart}, \citenamefont {Chen},\ and\ \citenamefont
  {Polzik}}]{vasilakis15}%
  \BibitemOpen
  \bibfield  {author} {\bibinfo {author} {\bibfnamefont {G.}~\bibnamefont
  {Vasilakis}}, \bibinfo {author} {\bibfnamefont {H.}~\bibnamefont {Shen}},
  \bibinfo {author} {\bibfnamefont {K.}~\bibnamefont {Jensen}}, \bibinfo
  {author} {\bibfnamefont {M.}~\bibnamefont {Balabas}}, \bibinfo {author}
  {\bibfnamefont {D.}~\bibnamefont {Salart}}, \bibinfo {author} {\bibfnamefont
  {B.}~\bibnamefont {Chen}}, \ and\ \bibinfo {author} {\bibfnamefont {E.~S.}\
  \bibnamefont {Polzik}},\ }\href@noop {} {\bibfield  {journal} {\bibinfo
  {journal} {Nat. Phys.}\ }\textbf {\bibinfo {volume} {11}},\ \bibinfo {pages}
  {389} (\bibinfo {year} {2015})}\BibitemShut {NoStop}%
\bibitem [{\citenamefont {Mhaskar}, \citenamefont {Knappe},\ and\ \citenamefont
  {Kitching}(2012)}]{mhaskar12}%
  \BibitemOpen
  \bibfield  {author} {\bibinfo {author} {\bibfnamefont {R.}~\bibnamefont
  {Mhaskar}}, \bibinfo {author} {\bibfnamefont {S.}~\bibnamefont {Knappe}}, \
  and\ \bibinfo {author} {\bibfnamefont {J.}~\bibnamefont {Kitching}},\
  }\href@noop {} {\bibfield  {journal} {\bibinfo  {journal} {Appl. Phys.
  Lett.}\ }\textbf {\bibinfo {volume} {101}},\ \bibinfo {pages} {241105}
  (\bibinfo {year} {2012})}\BibitemShut {NoStop}%
\bibitem [{\citenamefont {Sander}\ \emph {et~al.}(2012)\citenamefont {Sander},
  \citenamefont {Preusser}, \citenamefont {Mhaskar}, \citenamefont {Kitching},
  \citenamefont {Trahms},\ and\ \citenamefont {Knappe}}]{sander12}%
  \BibitemOpen
  \bibfield  {author} {\bibinfo {author} {\bibfnamefont {T.}~\bibnamefont
  {Sander}}, \bibinfo {author} {\bibfnamefont {J.}~\bibnamefont {Preusser}},
  \bibinfo {author} {\bibfnamefont {R.}~\bibnamefont {Mhaskar}}, \bibinfo
  {author} {\bibfnamefont {J.}~\bibnamefont {Kitching}}, \bibinfo {author}
  {\bibfnamefont {L.}~\bibnamefont {Trahms}}, \ and\ \bibinfo {author}
  {\bibfnamefont {S.}~\bibnamefont {Knappe}},\ }\href@noop {} {\bibfield
  {journal} {\bibinfo  {journal} {Biomed. Opt. Express}\ }\textbf {\bibinfo
  {volume} {3}},\ \bibinfo {pages} {981} (\bibinfo {year} {2012})}\BibitemShut
  {NoStop}%
\bibitem [{\citenamefont {Knappe}\ \emph {et~al.}(2010)\citenamefont {Knappe},
  \citenamefont {Sander}, \citenamefont {Kosch}, \citenamefont {Wiekhorst},
  \citenamefont {Kitching},\ and\ \citenamefont {Trahms}}]{knappe10}%
  \BibitemOpen
  \bibfield  {author} {\bibinfo {author} {\bibfnamefont {S.}~\bibnamefont
  {Knappe}}, \bibinfo {author} {\bibfnamefont {T.~H.}\ \bibnamefont {Sander}},
  \bibinfo {author} {\bibfnamefont {O.}~\bibnamefont {Kosch}}, \bibinfo
  {author} {\bibfnamefont {F.}~\bibnamefont {Wiekhorst}}, \bibinfo {author}
  {\bibfnamefont {J.}~\bibnamefont {Kitching}}, \ and\ \bibinfo {author}
  {\bibfnamefont {L.}~\bibnamefont {Trahms}},\ }\href@noop {} {\bibfield
  {journal} {\bibinfo  {journal} {Appl. Phys. Lett.}\ }\textbf {\bibinfo
  {volume} {97}},\ \bibinfo {pages} {133703} (\bibinfo {year}
  {2010})}\BibitemShut {NoStop}%
\bibitem [{\citenamefont {Schwindt}\ \emph {et~al.}(2004)\citenamefont
  {Schwindt}, \citenamefont {Knappe}, \citenamefont {Shah}, \citenamefont
  {Hollberg}, \citenamefont {Kitching}, \citenamefont {Liew},\ and\
  \citenamefont {Moreland}}]{schwindt04}%
  \BibitemOpen
  \bibfield  {author} {\bibinfo {author} {\bibfnamefont {P.~D.}\ \bibnamefont
  {Schwindt}}, \bibinfo {author} {\bibfnamefont {S.}~\bibnamefont {Knappe}},
  \bibinfo {author} {\bibfnamefont {V.}~\bibnamefont {Shah}}, \bibinfo {author}
  {\bibfnamefont {L.}~\bibnamefont {Hollberg}}, \bibinfo {author}
  {\bibfnamefont {J.}~\bibnamefont {Kitching}}, \bibinfo {author}
  {\bibfnamefont {L.-A.}\ \bibnamefont {Liew}}, \ and\ \bibinfo {author}
  {\bibfnamefont {J.}~\bibnamefont {Moreland}},\ }\href@noop {} {\bibfield
  {journal} {\bibinfo  {journal} {Appl. Phys. Lett.}\ }\textbf {\bibinfo
  {volume} {85}},\ \bibinfo {pages} {6409} (\bibinfo {year}
  {2004})}\BibitemShut {NoStop}%
\bibitem [{\citenamefont {Happer}\ and\ \citenamefont {Tang}(1973)}]{happer73}%
  \BibitemOpen
  \bibfield  {author} {\bibinfo {author} {\bibfnamefont {W.}~\bibnamefont
  {Happer}}\ and\ \bibinfo {author} {\bibfnamefont {H.}~\bibnamefont {Tang}},\
  }\href {\doibase 10.1103/PhysRevLett.31.273} {\bibfield  {journal} {\bibinfo
  {journal} {Phys. Rev. Lett.}\ }\textbf {\bibinfo {volume} {31}},\ \bibinfo
  {pages} {273} (\bibinfo {year} {1973})}\BibitemShut {NoStop}%
\bibitem [{\citenamefont {Allred}\ \emph {et~al.}(2002)\citenamefont {Allred},
  \citenamefont {Lyman}, \citenamefont {Kornack},\ and\ \citenamefont
  {Romalis}}]{allred02}%
  \BibitemOpen
  \bibfield  {author} {\bibinfo {author} {\bibfnamefont {J.~C.}\ \bibnamefont
  {Allred}}, \bibinfo {author} {\bibfnamefont {R.~N.}\ \bibnamefont {Lyman}},
  \bibinfo {author} {\bibfnamefont {T.~W.}\ \bibnamefont {Kornack}}, \ and\
  \bibinfo {author} {\bibfnamefont {M.~V.}\ \bibnamefont {Romalis}},\ }\href
  {\doibase 10.1103/PhysRevLett.89.130801} {\bibfield  {journal} {\bibinfo
  {journal} {Phys. Rev. Lett.}\ }\textbf {\bibinfo {volume} {89}},\ \bibinfo
  {pages} {130801} (\bibinfo {year} {2002})}\BibitemShut {NoStop}%
\bibitem [{\citenamefont {Cohen-Tannoudji}\ \emph {et~al.}(1970)\citenamefont
  {Cohen-Tannoudji}, \citenamefont {Dupont-Roc}, \citenamefont {Haroche},\ and\
  \citenamefont {Lalo{\"e}}}]{cohen70}%
  \BibitemOpen
  \bibfield  {author} {\bibinfo {author} {\bibfnamefont {C.}~\bibnamefont
  {Cohen-Tannoudji}}, \bibinfo {author} {\bibfnamefont {J.}~\bibnamefont
  {Dupont-Roc}}, \bibinfo {author} {\bibfnamefont {S.}~\bibnamefont {Haroche}},
  \ and\ \bibinfo {author} {\bibfnamefont {F.}~\bibnamefont {Lalo{\"e}}},\
  }\href@noop {} {\bibfield  {journal} {\bibinfo  {journal} {Rev. Phys. Appl.}\
  }\textbf {\bibinfo {volume} {5}},\ \bibinfo {pages} {95} (\bibinfo {year}
  {1970})}\BibitemShut {NoStop}%
\bibitem [{\citenamefont {Shah}\ and\ \citenamefont {Wakai}(2013)}]{shah13}%
  \BibitemOpen
  \bibfield  {author} {\bibinfo {author} {\bibfnamefont {V.~K.}\ \bibnamefont
  {Shah}}\ and\ \bibinfo {author} {\bibfnamefont {R.~T.}\ \bibnamefont
  {Wakai}},\ }\href@noop {} {\bibfield  {journal} {\bibinfo  {journal} {Phys.
  Med. Bio.}\ }\textbf {\bibinfo {volume} {58}},\ \bibinfo {pages} {8153}
  (\bibinfo {year} {2013})}\BibitemShut {NoStop}%
\bibitem [{\citenamefont {Johnson}, \citenamefont {Schwindt},\ and\
  \citenamefont {Weisend}(2010)}]{johnson10}%
  \BibitemOpen
  \bibfield  {author} {\bibinfo {author} {\bibfnamefont {C.}~\bibnamefont
  {Johnson}}, \bibinfo {author} {\bibfnamefont {P.~D.}\ \bibnamefont
  {Schwindt}}, \ and\ \bibinfo {author} {\bibfnamefont {M.}~\bibnamefont
  {Weisend}},\ }\href@noop {} {\bibfield  {journal} {\bibinfo  {journal} {Appl.
  Phys. Lett.}\ }\textbf {\bibinfo {volume} {97}},\ \bibinfo {pages} {243703}
  (\bibinfo {year} {2010})}\BibitemShut {NoStop}%
\bibitem [{\citenamefont {Preusser}\ \emph {et~al.}(2009)\citenamefont
  {Preusser}, \citenamefont {Knappe}, \citenamefont {Kitching},\ and\
  \citenamefont {Gerginov}}]{preusser09}%
  \BibitemOpen
  \bibfield  {author} {\bibinfo {author} {\bibfnamefont {J.}~\bibnamefont
  {Preusser}}, \bibinfo {author} {\bibfnamefont {S.}~\bibnamefont {Knappe}},
  \bibinfo {author} {\bibfnamefont {J.}~\bibnamefont {Kitching}}, \ and\
  \bibinfo {author} {\bibfnamefont {V.}~\bibnamefont {Gerginov}},\ }in\
  \href@noop {} {\emph {\bibinfo {booktitle} {2009 IEEE International Frequency
  Control Symposium Joint with the 22nd European Frequency and Time forum}}}\
  (\bibinfo {organization} {IEEE},\ \bibinfo {year} {2009})\ pp.\ \bibinfo
  {pages} {1180--1182}\BibitemShut {NoStop}%
\bibitem [{\citenamefont {Knappe}\ \emph {et~al.}(2016)\citenamefont {Knappe},
  \citenamefont {Alem}, \citenamefont {Sheng},\ and\ \citenamefont
  {Kitching}}]{knappe16}%
  \BibitemOpen
  \bibfield  {author} {\bibinfo {author} {\bibfnamefont {S.}~\bibnamefont
  {Knappe}}, \bibinfo {author} {\bibfnamefont {O.}~\bibnamefont {Alem}},
  \bibinfo {author} {\bibfnamefont {D.}~\bibnamefont {Sheng}}, \ and\ \bibinfo
  {author} {\bibfnamefont {J.}~\bibnamefont {Kitching}},\ }in\ \href@noop {}
  {\emph {\bibinfo {booktitle} {Journal of Physics: Conference Series}}},\
  Vol.\ \bibinfo {volume} {723}\ (\bibinfo {organization} {IOP Publishing},\
  \bibinfo {year} {2016})\ p.\ \bibinfo {pages} {012055}\BibitemShut {NoStop}%
\bibitem [{\citenamefont {Shah}\ and\ \citenamefont {Romalis}(2009)}]{shah09}%
  \BibitemOpen
  \bibfield  {author} {\bibinfo {author} {\bibfnamefont {V.}~\bibnamefont
  {Shah}}\ and\ \bibinfo {author} {\bibfnamefont {M.~V.}\ \bibnamefont
  {Romalis}},\ }\href {\doibase 10.1103/PhysRevA.80.013416} {\bibfield
  {journal} {\bibinfo  {journal} {Phys. Rev. A}\ }\textbf {\bibinfo {volume}
  {80}},\ \bibinfo {pages} {013416} (\bibinfo {year} {2009})}\BibitemShut
  {NoStop}%
\bibitem [{krz()}]{krzyzewskiprep}%
  \BibitemOpen
  \href@noop {} {}\bibinfo {note} {S. P. Krzyzewski, A. R. Perry, S. Geller, V.
  Gerginov, and S. Knappe, in preparation.}\BibitemShut {Stop}%
\bibitem [{Note1()}]{Note1}%
  \BibitemOpen
  \bibinfo {note} {Previously reported CMRR of SERF gradiometers is typically
  in the range of 10$\sim $100}\BibitemShut {NoStop}%
\bibitem [{\citenamefont {Seltzer}(2008)}]{seltzer08}%
  \BibitemOpen
  \bibfield  {author} {\bibinfo {author} {\bibfnamefont {S.~J.}\ \bibnamefont
  {Seltzer}},\ }\href@noop {} {Ph.D. thesis},\ \bibinfo  {school} {Princeton
  University} (\bibinfo {year} {2008})\BibitemShut {NoStop}%
\bibitem [{Note2()}]{Note2}%
  \BibitemOpen
  \bibinfo {note} {Trade name is stated for technical clarity and does not
  imply endorsement by NIST.}\BibitemShut {Stop}%
\bibitem [{\citenamefont {Seltzer}\ and\ \citenamefont
  {Romalis}(2004)}]{seltzer04}%
  \BibitemOpen
  \bibfield  {author} {\bibinfo {author} {\bibfnamefont {S.~J.}\ \bibnamefont
  {Seltzer}}\ and\ \bibinfo {author} {\bibfnamefont {M.~V.}\ \bibnamefont
  {Romalis}},\ }\href@noop {} {\bibfield  {journal} {\bibinfo  {journal} {Appl.
  Phys. Lett.}\ }\textbf {\bibinfo {volume} {85}},\ \bibinfo {pages} {4804}
  (\bibinfo {year} {2004})}\BibitemShut {NoStop}%
\bibitem [{\citenamefont {Belfi}\ \emph {et~al.}(2010)\citenamefont {Belfi},
  \citenamefont {Bevilacqua}, \citenamefont {Biancalana}, \citenamefont
  {Cecchi}, \citenamefont {Dancheva},\ and\ \citenamefont {Moi}}]{belfi10}%
  \BibitemOpen
  \bibfield  {author} {\bibinfo {author} {\bibfnamefont {J.}~\bibnamefont
  {Belfi}}, \bibinfo {author} {\bibfnamefont {G.}~\bibnamefont {Bevilacqua}},
  \bibinfo {author} {\bibfnamefont {V.}~\bibnamefont {Biancalana}}, \bibinfo
  {author} {\bibfnamefont {R.}~\bibnamefont {Cecchi}}, \bibinfo {author}
  {\bibfnamefont {Y.}~\bibnamefont {Dancheva}}, \ and\ \bibinfo {author}
  {\bibfnamefont {L.}~\bibnamefont {Moi}},\ }\href@noop {} {\bibfield
  {journal} {\bibinfo  {journal} {Rev. Sci. Instrum.}\ }\textbf {\bibinfo
  {volume} {81}},\ \bibinfo {pages} {065103} (\bibinfo {year}
  {2010})}\BibitemShut {NoStop}%
\bibitem [{Note3()}]{Note3}%
  \BibitemOpen
  \bibinfo {note} {We use a Thorlabs LDC 205C current supply for the large
  Helmholtz coil system, and a NIST home-built current source for the small
  compensation coil system. The noise of Thorlabs current supply is about one
  order of magnitude higher than that of the NIST home-built current
  source.}\BibitemShut {Stop}%
\bibitem [{\citenamefont {Torrey}(1956)}]{torrey56}%
  \BibitemOpen
  \bibfield  {author} {\bibinfo {author} {\bibfnamefont {H.~C.}\ \bibnamefont
  {Torrey}},\ }\href {\doibase 10.1103/PhysRev.104.563} {\bibfield  {journal}
  {\bibinfo  {journal} {Phys. Rev.}\ }\textbf {\bibinfo {volume} {104}},\
  \bibinfo {pages} {563} (\bibinfo {year} {1956})}\BibitemShut {NoStop}%
\bibitem [{\citenamefont {Bevilacqua}\ \emph {et~al.}(2016)\citenamefont
  {Bevilacqua}, \citenamefont {Biancalana}, \citenamefont {Chessa},\ and\
  \citenamefont {Dancheva}}]{bevilacqua16}%
  \BibitemOpen
  \bibfield  {author} {\bibinfo {author} {\bibfnamefont {G.}~\bibnamefont
  {Bevilacqua}}, \bibinfo {author} {\bibfnamefont {V.}~\bibnamefont
  {Biancalana}}, \bibinfo {author} {\bibfnamefont {P.}~\bibnamefont {Chessa}},
  \ and\ \bibinfo {author} {\bibfnamefont {Y.}~\bibnamefont {Dancheva}},\
  }\href@noop {} {\bibfield  {journal} {\bibinfo  {journal} {Appl. Phys. B}\
  }\textbf {\bibinfo {volume} {122}},\ \bibinfo {pages} {1} (\bibinfo {year}
  {2016})}\BibitemShut {NoStop}%
\end{thebibliography}
\end{document}